\documentclass[12pt,preprint]{aastex}

\newcommand{\be}{\begin{equation}}
\newcommand{\ee}{\end{equation}}

\slugcomment{Printed at \today}
\slugcomment{Accepted by ApJ Letter}
\shorttitle{Relation between Synchrotron luminosity and Doppler factor in Blazar and GRB} \shortauthors{Wu, Q., et al. }

\begin{document}

\title{A Uniform Correlation between Synchrotron Luminosity and Doppler Factor in Gamma-ray Bursts and Blazars: hint of similar intrinsic luminosities?}

\author{Qingwen Wu\altaffilmark{1}, Yuan-Chuan Zou\altaffilmark{1}, Xinwu Cao\altaffilmark{2}, Ding-Xiong Wang\altaffilmark{1}, and Liang Chen\altaffilmark{2}}

\altaffiltext{1}{School of Physics, Huazhong University of Science and Technology,
 Wuhan 430074, China; Email: qwwu@hust.edu.cn; zouyc@hust.edu.cn; dxwang@hust.edu.cn}

\altaffiltext{2}{Key Laboratory for Research in Galaxies and Cosmology,
 Shanghai Astronomical Observatory, Chinese Academy
of Sciences, Shanghai, 200030  China; Email: cxw@shao.ac.cn; chenliangew@hotmail.com}


\begin{abstract}
  We compile 23 Gamma-ray Bursts (GRBs) and 21 blazars with estimated Doppler factors, and the Doppler factors of GRBs
  are estimated from their Lorentz factors by assuming their jet viewing angles $\theta\rightarrow0^{\rm o}$.
  Using the conventional assumption that the prompt emission of GRBs is dominated by the synchrotron radiation,
  we calculate the synchrotron luminosity of GRBs from their total isotropic energy and burst duration. Intriguingly,
  we discover a uniform correlation between the synchrotron luminosity and Doppler factor, $L_{\rm syn}\propto \mathcal{D}^{3.1}$,
  for GRBs and blazars, which suggests that they may share some similar jet physics. One possible
  reason is that GRBs and blazars have, more or less, similar intrinsic synchrotron luminosities and
  both of them are strongly enhanced by the beaming effect. After Doppler and redshift-correction,
  we find that the intrinsic peak energy of the GRBs ranges from 0.1 to 3~keV with a typical value of 1 keV.
  We further correct the beaming effect for the observed luminosity of GRBs and find that there exists a positive
  correlation between the intrinsic synchrotron luminosity and peak energy for GRBs, which is
  similar to that of blazars. Our results suggest that both the intrinsic
  positive correlation and the beaming effect may be responsible for the observed tight correlation between
  the isotropic energy and the peak energy in GRBs (so called ``Amati" relation).

\end{abstract}


\section{Introduction}

  Blazars are a subclass of active galactic nuclei (AGNs), including
  flat-spectrum radio quasars (FSRQs) and BL Lac objects (BL Lacs), which are
  believed to be radio loud AGNs with their jets oriented at relatively small
  angles with respect to the line of sight. Special relativity plays a key
  role in spectral calculations of the relativistic plasma outflow. One of the most
  important effects is the relativistic boosting of the radiation, which
  is primarily governed by the value of the Doppler factor $\mathcal{D}=[\Gamma(1-\beta \cos \theta)]^{-1}$
  [$\Gamma=(1-\beta^2)^{-1/2}$ is bulk Lorentz factor, $\beta$ is jet speed
  divided by speed of light,  and $\theta$ is the angle between the jet and
  the line of sight]. Several different approaches have been proposed to estimate
  the Doppler factor $\mathcal{D}$ of the jets in AGNs. \citet{gh93} estimated
  the Doppler factor with the very long baseline interferometry (VLBI) core sizes/fluxes, and
  the X-ray fluxes assuming the X-ray emission to be produced
  by the synchrotron self-Compton (SSC) processes in the jets. \citet{la99} proposed
  that variability brightness temperature of the radio source may be caused by
  the relativistic jets, and the Doppler factor can be estimated by assuming the
  intrinsic brightness temperature of the source is limited to the equipartition
  value.
  The apparent jet speed $\beta_{\rm app}$ can be measured from the multi-epoch VLBI
  observations, with which the Lorentz factor and the viewing angle of the jet can be
  derived, if the Doppler factor $\mathcal{D}$ and $\beta_{\rm app}$ both correspond to the same underlying
  flow speed. \citet{ho09} calculated the Doppler factor, Lorentz factor and viewing angle
  for a sample of blazars from their variability brightness temperature and
  apparent jet speed, and found that the Lorentz factors of blazars range from 2 to 30 with
  a typical value $\Gamma\sim 15$.

  Gamma-ray bursts (GRBs) are the most luminous astrophysical events so far. It is
  well known that GRBs produce ultrarelativistic and beamed jets with total
  isotropic-equivalent energy $E_{\rm iso}\sim 10^{49-55}$ erg
  \citep[e.g.,][for reviews and the references therein]{ma06,zh11}. Unlike the large-scale
  jets observed in AGNs, the jets are not observed in GRBs directly. There are several
  methods to constrain the Lorentz factors of the prompt emission of the GRBs and the typical
  Lorentz factors are found to be around several hundreds, which are mainly
  based on the ``compactness" argument \citep[e.g.,][]{fe93,zou11}, the flux and temperature of the thermal component
  of some GRB spectra \citep[e.g.,][]{pe07}, or the early afterglow light curves with the signal of fireball
  deceleration \citep[e.g.,][]{sa99,li10,gh11}.

  The spectral energy distribution (SED) of blazars is characterized by two broad components:
  the low-energy component peaked at infrared-soft X-ray wavebands and the high-energy component
  peaked at MeV-GeV wavebands. There is unanimous consensus that the first broad peak is due
  to the synchrotron emission, while the nature of the second peak remains controversial.
  Comptonization is believed to be responsible for the high-energy $\gamma$-ray emission, which
  can be classified into two categories: the synchrotron self-Comptonization (SSC) model and the
  external Comptonization (EC) model, according to the origin of the soft seed photons. The
  external soft seed photons may originate from the accretion disks, the broad line regions and/or
  the dust tori \citep[e.g.,][for a review and the references therein]{bo07}. The nature of the prompt
  emission of GRBs is still a mystery. The SED of the prompt emission of GRBs is normally peaked at around
  several hundred keV to MeV waveband, which can be well fitted with a smoothly-jointed broken power-law,
  the so-called Band function \citep[e.g.,][]{ba93}. The most successful theory to explain the prompt emission
  of GRBs to date is nonthermal synchrotron emission caused by the internal shocks due to the collision of two
  relativistic shells \citep[e.g.,][]{ma94,ta96}, though it still suffers some criticisms \citep[e.g.,][for a recent review
  and the references therein]{zh11}. The synchrotron spectra may extend to a few hundred keV in the
  observer frame if the Lorentz factors of the GRBs $\Gamma\sim$100$-$1000, and the inverse Compton scattering
  of such photons leads to GeV and TeV spectral components \citep[e.g.,][]{ma94}.

  It is found that both blazars and GRBs seem to form a sequence respectively, as functions of
  the power, with their overall SEDs. However, the trends of these two sequences are opposite to each
  other. \citet{fo98} found that the peak frequency of synchrotron
  emission decreases with increasing
  luminosity for a sample containing $\sim$ 100 blazars  \citep[see also][]{gh98}. However,
  the peak energy of the $\nu F_{\nu}$ spectrum, $E_{\rm peak}$, was found positively correlated
  to the total isotropic energy \citep[$E_{\rm iso}$, e.g.,][]{am02,gg04} or the isotropic luminosity
  \citep[$L_{\rm iso}$, e.g.,][]{li04,yo04} in the GRBs.  A quite strong $E_{\rm iso}-E_{\rm peak}$
  correlation is also discovered for some individual GRBs, which has the similar slope and normalization as those defined by
  the different GRBs \citep[e.g.,][]{fi09,gh10,lu10}. The physical origin for these
  correlations (in particular, the opposite trends) in blazars and GRBs is still unclear. One possibility is that they are caused by the Doppler
  beaming effect, since that both the radiation and the peak frequency/energy are affected by the
  Doppler factor. \citet{ni08} and \citet{wu08} found that the negative correlation between the synchrotron
  peak frequency and luminosity in blazars becomes positive after correcting the Doppler boosting effect for
  these two quantities. \citet{liu10} proposed that the temporal and spectral
  evolution in GRB pulse can be explained with the Doppler beaming effect caused by the jet precession.

  In this \emph{Letter}, there are two motivations for studying the beaming effect in GRBs
  and blazars: (1) understand the intrinsic values of the peak energy and luminosity in GRBs after
  eliminating the Doppler boosting effect; (2) explore the possible similarities of the jet physics in
  GRBs and blazars. Throughout this work, we assume the following cosmology:
  $H_{0}=71\ \rm km\ s^{-1} Mpc^{-1}$, $\Omega_{0}=0.3$ and
  $\Omega_{\Lambda}=0.7$.

  \section{Sample}
   Our sample includes two parts: GRBs and blazars. The most important selection criterion is that
   their Doppler factors have been estimated in the literature or can be estimated in this work.
   We briefly describe the sample as follows (see also Table 1 for the details).

   The GRB sample includes 23 sources with redshift
   measurements, and the initial Lorentz factors of the prompt emission have been constrained from the early afterglow
   of optical and/or X-ray light curve with a signal of fireball deceleration, which are selected from \citet[][]{li10}.
   The Lorentz factor of GRB 080319C has been estimated from the optical light curve and X-ray light curve separately,
   and the Lorentz factors constrained with these two methods are included in our sample, as they are roughly consistent with each other.
   We note that the Lorentz factor estimated from the ``compactness" argument or blackbody component
   is not included in this work, due to it still suffering great uncertainties \citep[e.g.,][]{fe93,pe07}.
   The Doppler factors of GRBs are simply estimated with $\mathcal{D}\sim(1+\beta)\Gamma\sim2\Gamma$ due
   to the Lorentz factors of GRBs $\Gamma\gg1$ and the viewing angles $\theta\rightarrow0^{\rm o}$ \citep[e.g.,][]{ma06}.
   To compare with the blazars, we evaluate the luminosity of GRBs from their total isotropic energy
   $E_{\rm iso}$ and time duration $T_{\rm R45}$, where $E_{\rm iso}$ and $T_{\rm R45}$ are selected from \citet{bu07,bu10}.
   The total isotropic energy $E_{\rm iso}$ is defined in the energy between $1-10^{4}$ keV, and $T_{\rm R45}$ is the total time
   interval of the brightest bins in the light curve that contains 45\% of the burst
   fluence \citep{re01}. The peak energy of GRBs is a key parameter, which may
help us understand the jet radiation processes, and therefore,
   we also selected the data of the observed peak energy of GRBs from \citet[][]{bu07,bu10}.

   The blazar sample in this work consists of 21 sources. The Doppler factors, $\mathcal{D}$, of these blazars are estimated from the
   variability brightness temperature by assuming the intrinsic brightness temperature of the equipartition
   value \citep[see][for more details and the references therein]{sa10}. The blazars with Doppler factors estimated
   from the X-ray fluxes and VLBI core sizes/fluxes, assuming the SSC origin of
   X-ray emission, are not included in this work, since the hard X-ray emission of many blazars may be dominated by the external
   Compton emission \citep[e.g.,][]{gt10}. The second selection criterion of the blazars
   is the multi-frequency, simultaneous observational data of SEDs, which allows us to estimate their total
    synchrotron luminosities and the synchrotron peak energy.
   We select the blazars from \citet{ab10}, where all sources have quasi-simultaneous (within 3 months), broad-band observed SEDs.
   The peak frequency of synchrotron emission was estimated by
fitting the part of the SED dominated by the synchrotron
    emission with a third-degree polynomial equation \citep[see][for the details]{ab10}, which is also listed in Table 1.
   With the two criteria described above, 21 blazars are selected from \citet{sa10} and \citet{ab10}.

  \section{Results}
   \subsection{$L_{\rm syn}$-$\mathcal{D}$ Correlation in GRBs and Blazars}

   The Doppler factors of GRBs are estimated from their Lorentz factors with $\mathcal{D}\sim2 \Gamma$.
    The distributions of the Doppler factors of blazars and GRBs are presented in Figure 1. The average value
    $<$$\log \mathcal{D}$$>$=1.21 with standard deviation $\sigma=0.2$ for blazars, and $<$$\log \mathcal{D}$$>$=2.74 with
    $\sigma=0.25$ for GRBs respectively.

   Using the conventional assumption that the prompt emission of GRBs is dominated by the synchrotron emission,
   we calculate the isotropic synchrotron luminosity of each GRB from its total isotropic energy,
   $E_{\rm iso}$, and burst duration, $T_{\rm R45}$, with $L_{\rm syn}=E_{\rm iso}/\left[T_{\rm R45}(1+z)\right]$.
   We present the relation between the synchrotron luminosity, $L_{\rm syn}$, and the Doppler factor, $\mathcal{D}$,
   in Figure 2 (red circles). It is found that $L_{\rm syn}$ is positively correlated to $\mathcal{D}$ for these GRBs.
   The best fitting (long-dashed line) gives
   \be
   \log L_{\rm syn}=(3.38\pm0.50)\log \mathcal{D} + (42.85\pm1.36),
   \ee
   with the Spearman's rank correlation coefficient $r=0.86$ (chance probability $p<10^{-8}$).

   To estimate the total synchrotron emission of blazars, we integrate the best-fitted spectrum of the synchrotron component
   of every blazar
   \citep[see][for the details]{ab10}. The synchrotron luminosities
   of these balzars are listed in Table (1).  The correlation between $L_{\rm syn}$ and $\mathcal{D}$ for blazars is
   shown in Figure 2 (red squares). The best linear fitting (short-dashed line) gives
   \be
   \log L_{\rm syn}=(2.56\pm0.52)\log \mathcal{D} + (44.37\pm0.63),
   \ee
   with the correlation coefficient $r=0.67$ (chance probability $p=3.2\times10^{-4}$).

   Our results indicate that the synchrotron luminosity $L_{\rm syn}$ is closely correlated with $\mathcal{D}$ both for GRBs and blazars.
   Furthermore, it is found that GRBs and blazars roughly follow the same correlation between $L_{\rm syn}$ and
   $\mathcal{D}$ (see Figure 2). For the entire sample of GRBs and blazars, the best fitting (red solid line) gives
   \be
   \log L_{\rm syn}=(3.10\pm0.11)\log \mathcal{D} + (43.71\pm0.21).
   \ee
   The Spearman test gives the correlation coefficient $r=0.98$ and chance probability $p<10^{-8}$.
   The tight correlation between $L_{\rm syn}$ and $\mathcal{D}$ in both GRBs and blazars may provide
   a hint for their jet similarities, which will be discussed below.

 \subsection{The Intrinsic $E^{'}_{\rm peak}$-$L^{'}_{\rm syn}$ Correlation in GRBs and Blazars}

    The redshift and Doppler-correction of peak energy of GRBs and blazars can be calculated by
    using
   \be
   E^{'}_{\rm peak}=E_{\rm peak}\left[\frac{1+z}{\mathcal{D}} \right],
   \ee
  where the primed symbol represents the value measured in the source rest frame.
  We find that the values $E^{'}_{\rm peak}$ range from  $\sim 0.1$ to $\sim 3$ keV for GRBs, and
  $\sim 10^{-6}-10^{-3}$ keV for blazars, respectively.
    The distribution of $\log E^{'}_{\rm peak}$ is plotted in Figure 3,
  where $<$$\log E^{'}_{\rm peak}$$>=-$0.15 with standard deviation $\sigma=0.38$ for GRBs and $<$$\log E^{'}_{\rm peak}$$>=-$5.02 with
  $\sigma=0.48$ for blazars.



  The observed synchrotron emission being amplified due to the Doppler boosting effect can be described by
  $L_{\rm syn}=\mathcal{D}^{\alpha}L^{'}_{\rm syn}$ , where $L^{'}_{\rm syn}$ is the
  intrinsic synchrotron luminosity in the source rest frame. The beaming factor index $\alpha$ depends on the
  detailed physics of jets \citep[i.e.,  $\alpha=3$ for a continuous jet and $\alpha=4$ for a moving sphere, see][for more details]{gh93}.
  Here, we simply assume $\alpha=3.5$ (a middle value between 3 and 4) to estimate their intrinsic
  synchrotron luminosities.  The relation of the intrinsic synchrotron luminosity $L^{'}_{\rm syn}$ and the peak
  energy $E^{'}_{\rm peak}$ measured in the rest frame of the source is presented in Figure 4 for GRBs (squares)
  and blazars (circles). The Spearman's rank correlation coefficient $r=0.38$ (chance
  probability $p=0.05$) for the GRB sample, and $r=0.55$ (chance probability $p=0.005$) for the blazar sample,
  are derived.
  The Spearman test results suggest a positive correlation between the
  intrinsic quantities of $L^{'}_{\rm syn}$ and $E^{'}_{\rm peak}$ for GRBs and blazars respectively.
  The results are unchanged if we adopt a different value of $\alpha$ (i.e., $\alpha=3$ or $4$).

  \section{Discussion}


   The radiation from GRBs and blazars is believed to
   originate from the relativistic jets with very small viewing angles. Therefore, the beaming effect should be
   very important for both of these two kinds of sources. The strong correlation between the luminosity and Doppler factor
   in both GRBs and blazars indeed support this scenario. The most prominent result in this
   work is that both GRBs and blazars follow the same $L_{\rm syn}-\mathcal{D}$ correlation (see Figure 2),
   where $L_{\rm syn}$ is proportional to $\mathcal{D}^{3.1}$ for the whole sample. This relation should be
   very useful to investigate the jet physics in GRBs and blazars. One of the possibilities is that GRBs and blazars
   have, more or less, similar intrinsic luminosities, and their luminosities are enhanced by the beaming
   effect in the same way due to the possible similar jet physics (e.g., the beaming factor indices $\alpha$ is similar
   for both of them). We note that the slope of the best-fitted line of the blazar sample is shallower than that of the whole sample,
   while the best-fitted line of the GRB sample show a steeper slope. The physical reason is unclear, and more observational
   data are expected for the further test on this issue.
   The well-known definition of the $T_{90}$ duration is not adopted to calculate the luminosity
   due to the light curve of GRBs is normally very complex \citep[e.g.,][]{pa99}. In particular,
   there exist quiescent periods in the light curve of some long GRBs \citep[e.g.,][]{ra01}. Therefore, we choose the burst duration,
    $T_{\rm R45}$, in this work to estimate the luminosity of GRBs, even if we are not sure it is the best.

   The peak energy, $E_{\rm peak}$, is a crucial parameter in GRBs, at which most of the emission is radiated.
   After considering the redshift- and Doppler-corrections (equation 4), we find that
   the intrinsic values of the peak energy range from $\sim 0.1$ to $\sim3$ keV with a typical value $\sim 1$ keV for the
   GRBs in our sample, which is nearly five orders magnitude higher than that of blazars in our sample (see Figure 3).
   We note that the intrinsic peak energy of some GRBs with $E^{'}_{\rm peak}\sim0.1$ keV is just a little bit higher than
   that of \emph{high-frequency peaked} BL Lacs with peak energy of the synchrotron emission of $E^{'}_{\rm peak}\sim 0.01$ keV
   \citep[or $\nu^{'}_{\rm peak}\sim 10^{15}$ Hz, e.g.,][]{ni08}.

   The $E_{\rm peak}$-$E_{\rm iso}$ (or $E_{\rm peak}$-$L_{\rm iso}$) correlation is one of the most thoroughly studied correlations
   in GRBs. However, its physical origin is still unclear.  One possibility is that both peak energy and isotropic energy/luminosity
   are enhanced by the beaming effect. \citet{el06} proposed that the intrinsic peak energy and intrinsic isotropic energy are similar for
   different GRBs, and the observed positive correlation is caused by the beaming effect. We test this issue with our GRB sample.
   The intrinsic synchrotron
luminosity is derived with $L^{'}_{\rm syn}=L_{\rm
syn}/\mathcal{D}^{\alpha}$,
   where the beaming factor index $\alpha$ is 3 or 4 for a continuous jet or a moving sphere. Here we simply assume
   $\alpha=3.5$ due to the lack of enough knowledge of the jets. After eliminating the beaming effect for both quantities, we find that $E^{'}_{\rm peak}$ is
   still positively correlated to $L^{'}_{\rm syn}$ for GRBs, which is similar to that of blazars (see Figure 4). \citet{ni08} also pointed out that
   the blazar sequence may be the artefact of the Doppler boosting. They found that the negative correlation between the synchrotron peak
   frequency and the peak luminosities becomes positive after considering the Doppler corrections. The synchrotron peak luminosities
   equal approximately to the total synchrotron luminosities in our work.
   The similar trend in the relation of intrinsic peak energy
   and intrinsic luminosity suggests that both GRBs and blazars are probably governed by the similar jet physics.
   The physical mechanism for this positive correlation is still unclear, which may be partly
   caused by both of these two quantities depending on $B^{'}$ and $\gamma_{\rm p}$ (e.g., $E^{'}_{\rm peak}\propto B^{'}\gamma_{\rm p}^{2}$ and
   $L^{'}_{\rm syn}\propto B^{'2}\gamma_{\rm p}^{2}$), where $B^{'}$ is the magnetic field strength in comoving frame
   and $\gamma_{\rm p}$ is the energy of electrons emitting at the peaks of the SED. Therefore, our results indicate that both
   the beaming effect and the positive correlation between the intrinsic peak energy and intrinsic luminosity may be responsible for
   the observed tight correlation of $E_{\rm peak}$-$L_{\rm iso}$ (or $E_{\rm peak}$-$E_{\rm iso}$).

   The relations of $L_{\rm syn}-\mathcal{D}$ and $E^{'}_{\rm peak}$(or $\nu^{'}_{\rm peak}$)-$L^{'}_{\rm syn}$ in GRBs
   and blazars suggest that there may exist some physical similarities in their relativistic jets,
   besides their phenomenally similar viewing angle $\theta\rightarrow0^{\rm o}$ and relativistic jet seed ($v \rightarrow$
   light speed $c$). \citet{ww11} also found that the GRB afterglows share similar radiation processes with high frequency-peaked
   BL Lacs, which suggests that the jet emission in GRBs may be analogous to that of blazars. \citet{gh11} proposed that
   the beaming correction for the intrinsic luminosities of the GRBs may be different from that of blazars, where the beaming factor
   index $\alpha$ is 2 ($L^{'}_{\rm syn}=L_{\rm syn}/\mathcal{D}^{\alpha}$), not 3 or 4 as used in blazars,
   due to the emitting region in ``fireball" of GRBs have a radial distribution of velocities while the ``blob"
    of blazars have a mono-directional velocity. If this is the case, the correlation between the synchrotron
     luminosity and Doppler factor (equation 1) suggests that the intrinsic synchrotron luminosity of GRBs
     should be correlated with the Doppler factor. We find that our conclusion on the positive correlation
     between the intrinsic luminosites and the intrinsic peak energy of GRBs remains unchanged
     if we use the beaming factor index $\alpha=2$, through the intrinsic luminosity will be several orders of
     magnitude higher than that derived from the beaming factor index $\alpha=$ 3 or 4. It is still unclear whether the ``fireball"
     of GRBs is different from the ``blob" of blazars or not, since both of them may be caused by similar physical processes
    (e.g., internal shock scenario, Rees \& Meszaros 2002, or internal collision induced magnetic
    reconnection and turbulence model, Zhang \& Yan 2011, or Cannonball model, Dar \& de Rujula 2004, etc.).

    It should be noted that the assumption of synchrotron emission as the main radiation mechanism for the prompt emission of GRBs
    still suffers some criticisms \citep[e.g.,][]{zh11}. The prompt $\gamma$-ray emission can also be explained by other models
   (e.g., thermal emission from the photosphere, Pe'er \& Ryde 2011; or the inverse Compton emission of Cannonball model, Dar \& de Rujula 2004, etc.).
    If this is the case, our results of the $E^{'}_{\rm peak}$ distribution and
    $E^{'}_{\rm peak}$-$L^{'}_{\rm syn}$ correlation in the comoving frame for GRBs also provide useful clues to constrain these models.
     More detailed analysis is required to further constrain these models, which will be our future work.

 \section*{Acknowledgments}

This work is supported by the NSFC (grants 11143001, 11103003, 11133005, 11173011, 10821302, 10833002,
and 10873005), the National Basic Research Program of China (2009CB824800),
the Science and Technology Commission of Shanghai Municipality (10XD1405000),
 the Doctoral Program of Higher Education (200804870050), the HUST (01-24-012030), and
 the Fundamental Research Funds for the Central Universities (HUST: 2011TS159).

\begin{table}[t]
\footnotesize
  \centerline{\bf Table 1. The data of GRB and Blazar}
  \begin{tabular}{lcccccccc}\hline
Source & $z$ & $\log E_{\rm iso}$ & $T_{\rm R45}$ & $\log L_{\rm syn}$& $\Gamma_{0}$  & $\theta$ & $\mathcal{D}$
& $E_{\rm peak}$\\
         &        & ($10^{52}\ \rm ergs$) & (s) & ($\rm ergs\ s^{-1}$) &   &   &   &  (keV)\\
\hline
          &        &                         &                &             GRB              &                &           &                    &                      \\
050730   &  3.97  & $    9^{+8}_{-3}      $ &   $21\pm1    $ &  $52.33^{+0.28}_{-0.18}$ &  $289^{+41}_{-28}$  & $\sim 0^{\rm o}$  & $ 578^{+82}_{-56}$   &   $196^{+563}_{-87}  $ \\
050820A  &  2.615 & $   97^{+31}_{-14}    $ &   $11\pm0.8  $ &  $53.50^{+0.12}_{-0.07}$ &  $332^{+42}_{-21}$  & $\sim 0^{\rm o}$  & $ 664^{+84}_{-42}$   &   $490^{+720}_{-300} $ \\
060418   &  1.49  & $   10^{+7}_{-2}      $ &   $14.5\pm0.5$ &  $52.23^{+0.23}_{-0.10}$ &  $379^{+33}_{-10}$  & $\sim 0^{\rm o}$  & $ 758^{+66}_{-20}$   &   $217^{+472}_{-87}  $ \\
060605   &  3.8   & $  2.5^{+3.1}_{-0.6}  $ &   $5.4\pm0.5 $ &  $52.35^{+0.35}_{-0.12}$ &  $283^{+44}_{-9 }$  & $\sim 0^{\rm o}$  & $ 566^{+88}_{-18}$   &   $142^{+359}_{-50}  $ \\
060607A  &  3.082 & $    9^{+7}_{-2}      $ &   $15.1\pm0.8$ &  $52.39^{+0.25}_{-0.11}$ &  $426^{+41}_{-12}$  & $\sim 0^{\rm o}$  & $ 852^{+82}_{-24}$   &   $139^{+218}_{-41}  $ \\
060904B  &  0.703 & $ 0.72^{+0.43}_{-0.43}$ &   $5.9\pm0.6 $ &  $51.32^{+0.20}_{-0.39}$ &  $155^{+14}_{-14}$  & $\sim 0^{\rm o}$  & $ 310^{+28}_{-28}$   &   $ 83^{+128}_{-25}  $ \\
061007   &  1.262 & $104.7^{+6.9}_{-6.9}  $ &   $16.8\pm0.2$ &  $53.15^{+0.03}_{-0.03}$ &  $627^{+5 }_{-5 }$  & $\sim 0^{\rm o}$  & $1254^{+10}_{-10}$   &   $840^{+1100}_{-330}$ \\
070318   &  0.84  & $ 1.45^{+0.38}_{-0.38}$ &   $10.3\pm0.5$ &  $51.41^{+0.10}_{-0.13}$ &  $206^{+10}_{-10}$  & $\sim 0^{\rm o}$  & $ 412^{+20}_{-20}$   &   $196^{+445}_{-78}  $ \\
070411   &  2.954 & $   10^{+8}_{-2}      $ &   $32\pm1    $ &  $52.09^{+0.26}_{-0.10}$ &  $299^{+30}_{-8} $  & $\sim 0^{\rm o}$  & $ 598^{+60}_{-16}$   &   $120^{+556}_{-39}  $ \\
070419   &  0.97  & $ 0.24^{+0.23}_{-0.05}$ &   $37\pm4    $ &  $50.11^{+0.29}_{-0.10}$ &  $131^{+16}_{-4} $  & $\sim 0^{\rm o}$  & $ 262^{+32}_{ -8}$   &   $ 27^{+16}_{-19}   $ \\
071010   &  0.98  & $ 0.13^{+0.24}_{-0.01}$ &   $4.7\pm0.1 $ &  $50.74^{+0.45}_{-0.03}$ &  $145^{+34}_{-4} $  & $\sim 0^{\rm o}$  & $ 290^{+68}_{ -8}$   &   $ 37^{+49}_{-35}   $ \\
071031   &  2.692 & $  3.9^{+4.1}_{-0.6}  $ &   $34\pm4    $ &  $51.63^{+0.31}_{-0.07}$ &  $191^{+25}_{-4} $  & $\sim 0^{\rm o}$  & $ 382^{+50}_{ -8}$   &   $ 12^{+6}_{-11}    $ \\
080319C  &  1.95  & $ 22.6^{+3.4}_{-3.4}  $ &   $5.0\pm0.3 $ &  $53.12^{+0.06}_{-0.07}$ &  $327^{+7 }_{-7} $  & $\sim 0^{\rm o}$  & $ 654^{+14}_{-14}$   &   $157^{+303}_{-50}  $ \\
080330   &  1.51  & $ 0.41^{+0.94}_{-0.06}$ &   $4.0\pm0.6 $ &  $51.41^{+0.52}_{-0.07}$ &  $150^{+43}_{-3} $  & $\sim 0^{\rm o}$  & $ 300^{+86}_{ -6}$   &   $ 20^{+6}_{-19}    $ \\
080710   &  0.845 & $  0.8^{+0.8}_{-0.4}  $ &   $23\pm3    $ &  $50.81^{+0.30}_{-0.30}$ &  $90 ^{+11}_{-6} $  & $\sim 0^{\rm o}$  & $ 180^{+22}_{-12}$   &   $300^{+550}_{-200} $ \\
080810   &  3.35  & $   30^{+20}_{-20}    $ &   $31\pm2    $ &  $52.62^{+0.22}_{-0.48}$ &  $588^{+49}_{-49}$  & $\sim 0^{\rm o}$  & $1176^{+98}_{-98}$   &   $370^{+620}_{-220} $ \\
081203A  &  2.1   & $   17^{+13}_{-4}     $ &   $31\pm1    $ &  $52.23^{+0.25}_{-0.12}$ &  $315^{+30}_{-9} $  & $\sim 0^{\rm o}$  & $ 630^{+60}_{-18}$   &   $201^{+440}_{-75}  $ \\
070208$^a$   &  1.165 & $ 0.28^{+0.22}_{-0.08}$ &   $4.6\pm0.8 $ &  $51.12^{+0.25}_{-0.15}$ &  $115^{+23}_{-20}$  & $\sim 0^{\rm o}$  & $ 230^{+46}_{-40}$   &   $ 66^{+179}_{-33}  $ \\
080319C$^a$  &  1.95  & $ 22.6^{+3.4}_{-3.4}  $ &   $5.0\pm0.3$  &  $53.12^{+0.06}_{-0.07}$ &  $301^{+39}_{-39}$  & $\sim 0^{\rm o}$  & $ 602^{+78}_{-78}$   &   $157^{+303}_{-50}  $ \\
991203   &  1.60  & $436.5^{+60}_{-60}    $ &   $  ...     $ &  $...                  $ &  $\sim 966      $   & $\sim 0^{\rm o}$  & $   \sim 1932  $     &   $781^{+62}_{-62}   $ \\
050922C  &  2.198 & $ 5.06^{+0.55}_{-0.55}$ &   $1.2\pm0.04$ &  $53.13^{+0.04}_{-0.05}$ &  $\sim 401      $   & $\sim 0^{\rm o}$  & $   \sim 802   $     &   $183^{+267}_{-55}  $ \\
060210   &  3.91  & $ 41.5^{+5.7}_{-5.7}  $ &   $36\pm2    $ &  $52.75^{+0.06}_{-0.06}$ &  $\sim 381      $   & $\sim 0^{\rm o}$  & $   \sim 762   $     &   $136^{+347}_{-39}  $ \\
071010B  &  0.947 & $ 2.55^{+0.41}_{-0.41}$ &   $4.7\pm0.1 $ &  $52.02^{+0.06}_{-0.08}$ &  $\sim 309      $   & $\sim 0^{\rm o}$  & $   \sim 618   $     &   $ 56^{+8}_{-8}     $ \\
071112C  &  0.822 & $ 1.79^{+0.26}_{-0.26}$ &     ...        &      ...                 &  $\sim 244      $   & $\sim 0^{\rm o}$  & $   \sim 488   $     &   $422^{+137}_{-87}  $ \\
         &        &                         &                &            blazar        &                 &                   &                      &                      \\
0133+476 &  0.859 &      ...                &    ...         &  47.92                   &    14.4         &   $2.5^{\rm o}$      &   20.5             &   $10^{-3.78}$        \\
0234+285 &  1.207 &      ...                &    ...         &  47.71                   &    12.7         &   $3.5^{\rm o}$      &   16.0             &   $10^{-4.58}$        \\
0420+014 &  0.915 &      ...                &    ...         &  47.66                   &    11.3         &   $1.9^{\rm o}$      &   19.1             &   $10^{-3.98}$        \\
0528+134 &  2.07  &      ...                &    ...         &  48.06                   &    21.4         &   $1.7^{\rm o}$      &   30.9             &   $10^{-4.58}$        \\
0716+714 &  0.31  &      ...                &    ...         &  47.48                   &    10.2         &   $5.3^{\rm o}$      &   10.8             &   $10^{-2.78}$        \\
0851+202 &  0.306 &      ...                &    ...         &  47.32                   &    9.3          &   $1.9^{\rm o}$      &   16.8             &   $10^{-3.98}$        \\
1055+018 &  0.888 &      ...                &    ...         &  48.03                   &    8.8          &   $4.4^{\rm o}$      &   12.1             &   $10^{-4.28}$        \\
1156+295 &  0.729 &      ...                &    ...         &  47.40                   &    25.1         &   $2.0^{\rm o}$      &   28.2             &   $10^{-4.28}$        \\
1226+023 &  0.158 &      ...                &    ...         &  46.98                   &    13.8         &   $3.3^{\rm o}$      &   16.8             &   $10^{-3.88}$        \\
1253+055 &  0.536 &      ...                &    ...         &  47.54                   &    20.8         &   $2.4^{\rm o}$      &   23.8             &   $10^{-4.78}$        \\
1308+326 &  0.997 &      ...                &    ...         &  47.87                   &    22.0         &   $3.6^{\rm o}$      &   15.3             &   $10^{-4.28}$        \\
1502+106 &  1.839 &      ...                &    ...         &  48.39                   &    15.2         &   $4.7^{\rm o}$      &   11.9             &   $10^{-3.78}$        \\
1510+089 &  0.36  &      ...                &    ...         &  46.53                   &    20.6         &   $3.4^{\rm o}$      &   16.5             &   $10^{-4.28}$        \\
1749+096 &  0.322 &      ...                &    ...         &  46.47                   &    8.0          &   $4.2^{\rm o}$      &   11.9             &   $10^{-4.28}$        \\
2200+420 &  0.069 &      ...                &    ...         &  45.89                   &    5.4          &   $7.5^{\rm o}$      &   7.2              &   $10^{-3.78}$        \\
2251+158 &  0.859 &      ...                &    ...         &  48.46                   &    19.5         &   $1.3^{\rm o}$      &   32.9             &   $10^{-3.78}$        \\
0814+425 &  0.245 &      ...                &    ...         &  45.66                   &    2.7          &   $8.6^{\rm o}$      &   4.6              &   $10^{-4.18}$        \\
1803+784 &  1.814 &      ...                &    ...         &  47.23                   &    9.4          &   $4.5^{\rm o}$      &   12.1             &   $10^{-3.88}$        \\
2227+088 &  0.684 &      ...                &    ...         &  48.13                   &    10.0         &   $3.0^{\rm o}$      &   15.8             &   $10^{-3.88}$        \\
2230+114 &  1.037 &      ...                &    ...         &  47.66                   &    15.4         &   $3.7^{\rm o}$      &   15.5             &   $10^{-4.18}$        \\
0235+164 &  0.94  &      ...                &    ...         &  47.83                   &    12.1         &   $0.4^{\rm o}$      &   24.0             &   $10^{-3.88}$        \\

\hline
\end{tabular}

\begin{minipage}{170mm}
$^a$ The Lorentz factors are estimated from their X-ray light
curves, while they are estimated from the optical light curves for
all other GRBs.

\end{minipage}

\end{table}


\clearpage

\begin{figure}
\epsscale{1.0} \plotone{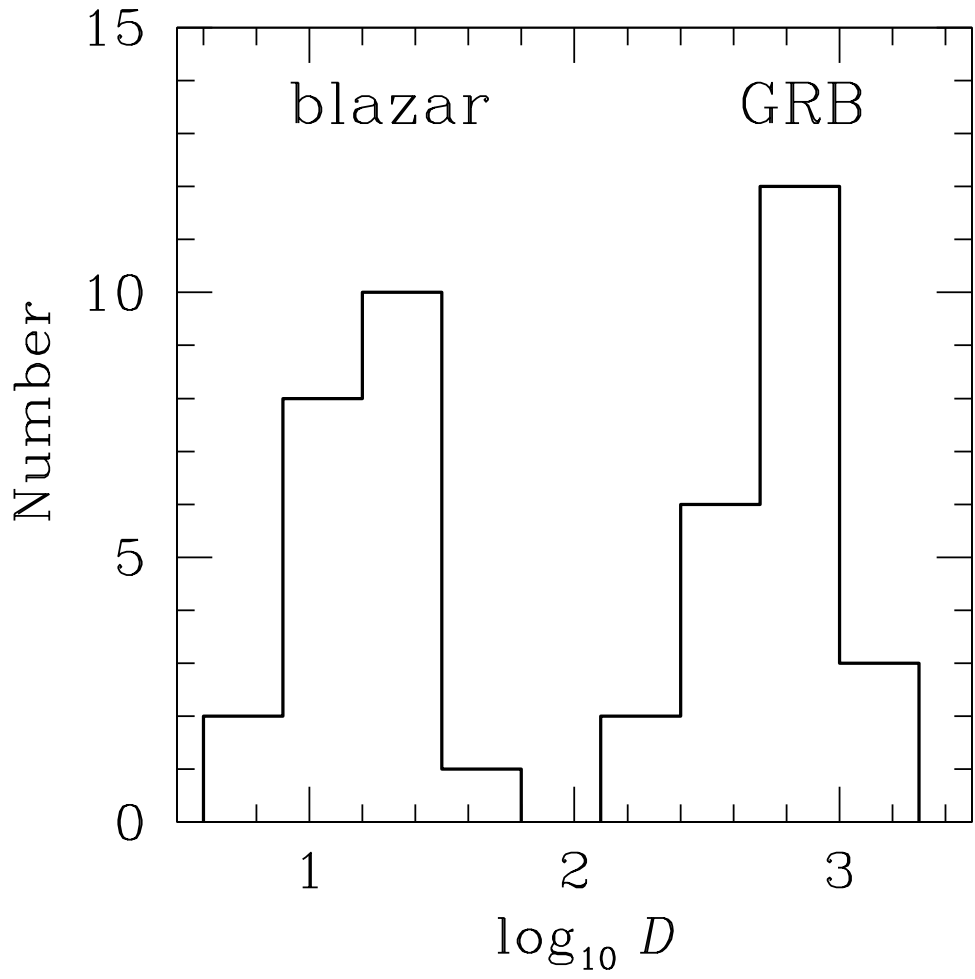} \caption{Distribution of the Doppler
factors, $\mathcal{D}$,
 for blazars and GRBs. \label{fig1}}
\end{figure}

\begin{figure}
\epsscale{1.0} \plotone{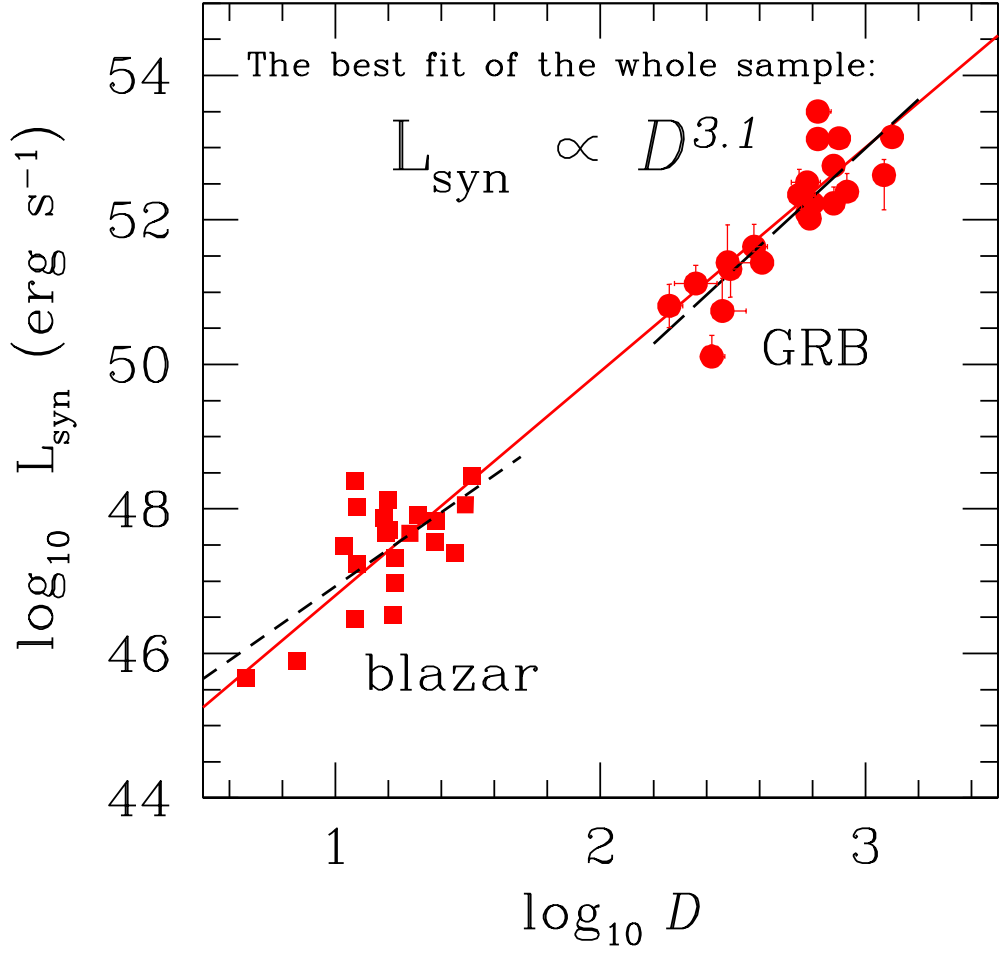} \caption{ The synchrotron
luminosity, $L_{\rm syn}$, vs Doppler factor, $\mathcal{D}$, correlation for GRBs and blazars. The long-dashed
line, short-dashed line and solid line represent the best linear fits of GRBs, blazars and both of them
respectively.
  \label{fig1}}
\end{figure}

\begin{figure}
\epsscale{1.0} \plotone{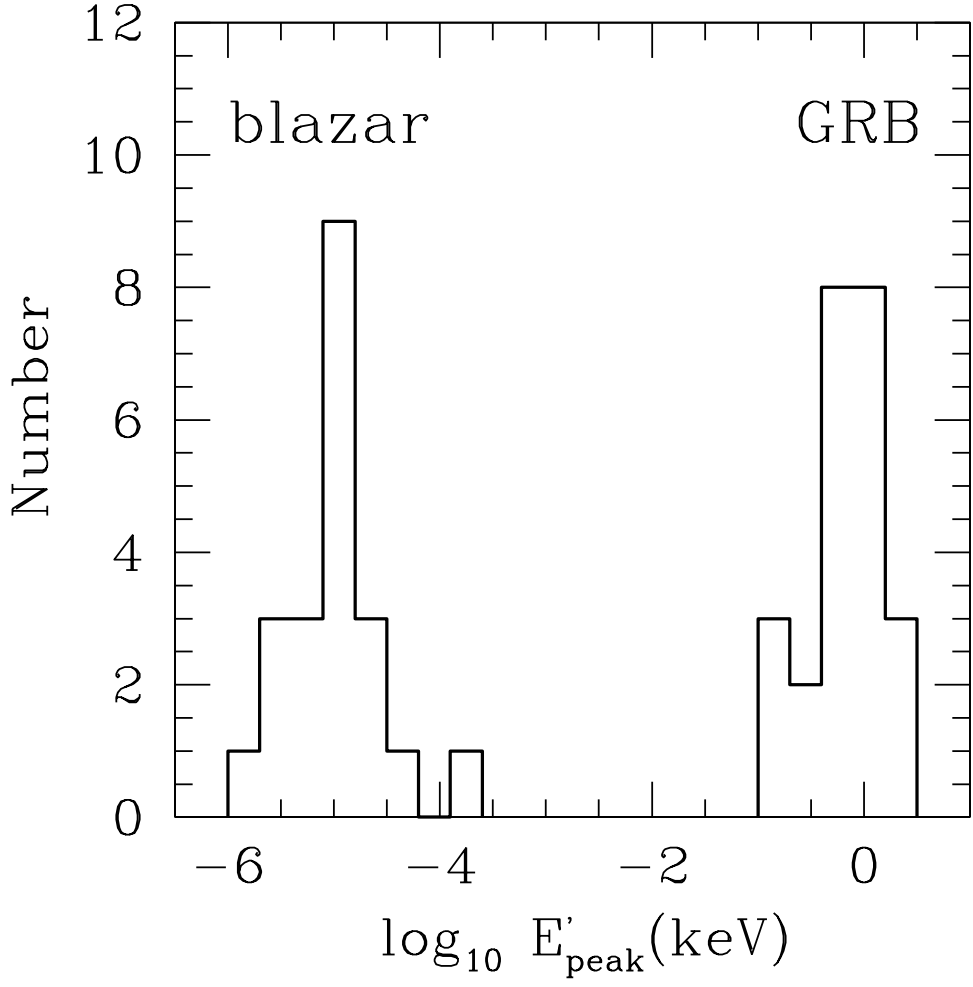} \caption{Distribution of the intrinsic peak energy, $E^{'}_{\rm peak}$,
 for blazars and GRBs after redshift- and Doppler-correction. \label{fig2}}
\end{figure}

\begin{figure}
\epsscale{1.0} \plotone{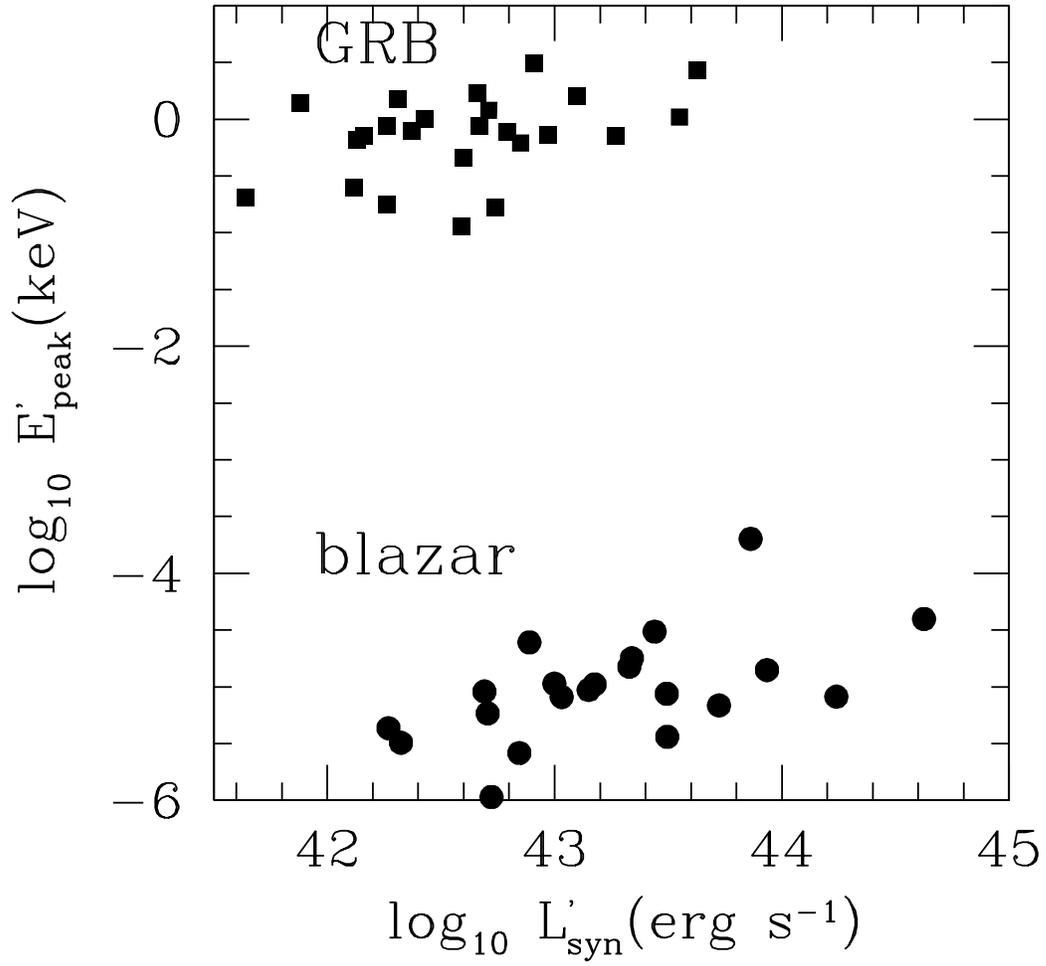} \caption{Correlation between the intrinsic
peak energy, $E^{'}_{\rm peak}$, and the intrinsic synchrotron luminosity $L^{'}_{\rm syn}$ after
correcting the beaming effect with $L^{'}_{\rm syn}$=$L_{\rm syn}/\mathcal{D}^{3.5}$.
 \label{fig3}}
\end{figure}


\begin{thebibliography}{}


\bibitem[Abdo et al.(2010)]{ab10}
Abdo, A. A., et al. 2010, \apj, 716, 30

\bibitem[Amati et al.(2002)]{am02}
Amati, L., et al. 2002, \aap, 390, 81

\bibitem[Band et al.(1993)]{ba93}
Band, D., et al. 1993, \apj, 413, 281

\bibitem[$\rm B\ddot{o}ttcher$(2007)]{bo07}
$\rm B\ddot{o}ttcher$, M. 2007, Ap\&SS, 309, 95

\bibitem[Butler et al.(2010)]{bu10}
Butler, N. R., Bloom, J. S., Poznanski, D. 2010, \apj, 711, 495

\bibitem[Butler et al.(2007)]{bu07}
Butler, N. R., Kocevski, D., Bloom, J. S., Curtis, J. L. 2007, \apj, 671, 656

\bibitem[Dar \& de Rujula(2004)]{dd04}
Dar, A., \& de Rujula, A. 2004, Physics Reports, 405, 203

\bibitem[Eichler \& Levinson(2006)]{el06}
Eichler, D., \& Levinson, A. 2006, \apj, 649, 5

\bibitem[Fenimore et al.(1993)]{fe93}
Fenimore, E. E., Epstein, R. I., \& Ho, C. 1993, A\&AS, 97, 59

\bibitem[Firmani et al.(2009)]{fi09}
Firmani, C., Cabrera, J. I., Avila-Reese, V., Ghisellini, G.,
Ghirlanda, G., Nava, L., \& Bosnjak, Z. 2009, 393, 1209

\bibitem[Fossati et al.(1998)]{fo98}
Fossati, G., Maraschi, L., Celotti, A., Comastri, A., Ghisellini, G. 1998, \mnras, 299, 433

\bibitem[Ghirlanda et al.(2004)]{gg04}
Ghirlanda, G., Ghisellini, G., \& Lazzati, D. 2004, \apj, 616, 331

\bibitem[Ghirlanda et al.(2010)]{gh10}
Ghirlanda, G., Nava, L., \& Ghisellini, G. 2010, \aap, 511, 43

\bibitem[Ghirlanda et al.(2011)]{gh11}
Ghirlanda, G., Nava, L., Ghisellini, G., Celotti, A., Burlon, D., Covino, S., \& Melandri, A. 2011, submitted to \mnras

\bibitem[Ghisellini et al.(2010)]{gt10}
Ghisellini, G., Tavecchio, F., Foschini, L., Ghirlanda, G., Maraschi, L., \& Celotti, A. 2010, \mnras, 402, 497

\bibitem[Ghisellini et al.(1998)]{gh98}
Ghisellini, G., Celotti, A., Fossati, G., Maraschi, L., \& Comastri, A. 1998, \mnras, 301, 451

\bibitem[Ghisellini et al.(1993)]{gh93}
Ghisellini, G.; Padovani, P.; Celotti, A.; \& Maraschi, L. 1993, \apj, 407, 65

\bibitem[Hovatta et al.(2009)]{ho09}
Hovatta, T., Valtaoja, E., Tornikoski, M., \& $\rm L\ddot{a}hteenm\ddot{a}ki$, A. 2009, \aap, 494, 297

\bibitem[$\rm L\ddot{a}hteenm\ddot{a}ki$ \& Valtaoja(1999)]{la99}
$\rm L\ddot{a}hteenm\ddot{a}ki$, A., \& Valtaoja, E. 1999, \apj, 521, 493

\bibitem[Liang et al.(2004)]{li04}
Liang, E. W., Dai, Z. G., \& Wu, X. F. 2004, \apj, 606, 29

\bibitem[Liang et al.(2010)]{li10}
Liang, E.-W., Yi, S.-X., Zhang, J., $\rm L\ddot{u}$, H.-J., Zhang, B.-B., Zhang, B. 2010, \apj, 725, 2209

\bibitem[Liu et al.(2010)]{liu10}
 Liu, T., Liang, E.-W., Gu, W.-M., Zhao, X.-H., Dai, Z.-G., \& Lu, J.-F. 2010, \aap, 516, 16

\bibitem[Lu et al.(2010)]{lu10}
Lu, R.-J., Hou, S.-J., Liang, E.-W. 2010, \apj, 720, 1146

\bibitem[$\rm M \acute{e} sz \acute{a} ros$(2006)]{ma06}
$\rm M \acute{e} sz \acute{a} ros$, P. 2006, Reports on Progress in Physics, 69, 2259


\bibitem[$\rm M \acute{e} sz \acute{a} ros$ et al.(1994)]{ma94}
$\rm M \acute{e} sz \acute{a} ros$, P., Rees, M. J., \& Papathanassiou, H. 1994, \apj, 432, 181

\bibitem[Nieppola et al.(2008)]{ni08}
Nieppola, E., Valtaoja, E., Tornikoski, M., Hovatta, T., \& Kotiranta, M. 2008, \aap, 488, 867

\bibitem[Paciesas et al.(1999)]{pa99}
Paciesas, W. S., et al. 1999, \apjs, 122, 465

\bibitem[Pe'er \& Ryde(2011)]{pr11}
Pe'er, A., \& Ryde, F. 2011, \apj, 732, 49

\bibitem[Pe'er et al.(2007)]{pe07}
Pe'er, A., Ryde, F., Wijers, R. A. M. J., $\rm M \acute{e} sz \acute{a} ros$, P., \& Rees, M. J. 2007, 664, 1

\bibitem[Ramirez-Ruiz et al.(2001)]{ra01}
Ramirez-Ruiz, E., Merloni, A., \& Rees, M. J. 2001, 324, 1147

\bibitem[Rees \& Meszaros(1994)]{rm94}
Rees, M. J., \& Meszaros, P. 1994, \apj, 430, 93

\bibitem[Rees(1978)]{re78}
Rees, M. J. 1978, \mnras, 184, 61


\bibitem[Reichart et al.(2001)]{re01}
Reichart, D. E., Lamb, D. Q., Fenimore, E. E., Ramirez-Ruiz, E., Cline, T. L., \& Hurley, K. 2001, \apj, 552, 57

\bibitem[Ryde(2004)]{ry04}
Ryde, F. 2004, \apj, 614, 827

\bibitem[Sari \& Piran(1999)]{sa99}
Sari, R., \& Piran, T. 1999, 520, 641

\bibitem[Savolainen et al.(2010)]{sa10}
Savolainen, T., Homan, D. C., Hovatta, T., Kadler, M., Kovalev, Y. Y., Lister, M. L., Ros, E., \& Zensus, J. A.
2010, \aap, 512, 24

\bibitem[Tavani(1996)]{ta96}
Tavani, M. 1996, \apj, 466, 768

\bibitem[Yonetoku et al.(2004)]{yo04}
Yonetoku, D., Murakami, T., Nakamura, T., Yamazaki, R., Inoue, A. K., Ioka, K. 2004, \apj, 609, 935

\bibitem[Wang \& Wei(2011)]{ww11}
Wang, J., \& Wei, J. Y. 2011, \apj, 726, 4

\bibitem[Wu et al.(2008)]{wu08}
Wu, Z.-Z., Gu, M.-F., \& Jiang, D.-R. 2008, Research in Astronomy and Astrophysics, 9, 168

\bibitem[Zhang \& $\rm M \acute{e} sz \acute{a} ros$(2002)]{zm02}
Zhang, B., \& $\rm M \acute{e} sz \acute{a} ros$, P. 2002, \apj, 581, 1236

\bibitem[Zhang \& Yan(2011)]{zh11}
 Zhang, B., \& Yan, H. 2011, \apj, 726, 90

\bibitem[Zhang(2011)]{zh11}
 Zhang, B. 2011, To appear in a special issue of Comptes Rendus Physique "GRB studies in the SVOM era",
 Eds. F. Daigne, G. Dubus, arXiv: 1104.0932

\bibitem[Zou et al.(2011)]{zou11}
 Zou, Y.-C., Fan, Y.-Z., Piran, T. 2011, \apj, 726, 2


\end{thebibliography}
\end{document}